# Anomalous Electromagnetics of Carbon Condensates in the Light of Ideas of a High-Temperature Superconductivity

S.G.Lebedev

Institute for Nuclear Research of Russian Academy of Sciences

**Abstract.**
The anomalies of electromagnetic properties of carbon condensates are discussed. Basic attention has been done to electromagnetic properties of thin carbon films produced by means of sputtering of spectral pure graphite in electrical arc (CA films). The room temperature rf-to-dc conversion associated with the ac Josephson effect, the jumps of conductivity stipulated by variation of current and temperature and zero spin density can be pointed among the anomalous properties of CA films. The number of basic hypotheses are presented in trying to explain this anomalies. As the main hypothesis the high temperature superconductivity is considered. As alternative the analogy with the chalcogenide glasses, fluctuation conductivity, etc. can be considered.

**1. Introduction**

In the given activity the attempt of explanation of anomalous conductive properties of some carbon condensates is undertaken within the framework of ideas and approaches of a high-temperature superconductivity (HTS). The given attempt is not the first and, probably, not the last.

First of all it is possible to mention cases assembled in a collection of a "room" and not reproduced superconductivity in the review /1/.

The other direction of "room" superconductivity is marked by research of polymeric films. The activities /2-4/ may give the representation about hard path of the authors on the research and "popularization" of a possible superconducting condition in polymeric films and capillary tubes at room temperatures and higher, up to *700K*. The authors do not lose hope, that the researched system represents direct implementation of Little's idea /5/ about HTS with the critical temperature *Tc > 300K*. In spite of the fact that the given direction is developed rather intensively (about 50 publications on the present time) the problem of a nature of a possible superconducting condition until now is not solved, probably, because of a complex spatial structure of conductive channels.

The third direction of "room" HTS can conditionally be designated as "carbon" HTS. The given direction in the instrumental plan continues the perovskit's line because in both cases the structure is granular and a basic means of research of such structure is the Josephson's effect. In the activity /6/ the possible HTS phase in a carbon - copper mixture was studied. The powder mixture of two components ($C_{60}$: *Cu = 7:1*) after an annealing was subjected to mechanical compression, and the value of the electroresistance *R* of this mixture depend on a degree of compression. It is reported that at *R < 200 Ohm* the sharp jump was observed in a temperature curve *R (T)* and at *T < 180K* the resistance start to be the instrumental zero.

In the past few years much attention has been payed to HTS signs in the carbon single-walled nanotubes (SWNT) and multiwalled nanotubes (MWNT) /7-8/. Such kind of object consits of cocentric graphite sheets which are curved in a long cylinder.

Characteristics of the specified measurements are that all of them were conducted on an alternating current with the use of Josephson's effect. Generally, as it is represented, the Josephson's effect plays a key role at research of superconductivity in inhomogeneous granular samples. This is because that in many cases in granular systems the resistance between granules is so high, that the system cannot gain a condition of common phase coherence at final temperatures. However phase coherence is present inside the separate granules or in the domain consist of final their number. The reverse Joserphson's effect (RJE) consists in detecting of a microwave radiation by system of superconducting granules. The interesting application of RJE at the beginning of a "room" superconductivity is circumscribed in activity /9/. The two-phase HTS ceramic *Y-Ba-Cu-O* was studied. One of the phase had the critical temperature of superconducting transition about 90 K, and the second, as it was supposed, had a beginning of superconducting transition at the *Tc = 240K*. In the specified activity the experimental technique is circumscribed which allows to separate RJE from other ways of a rectification of an alternating current. The behavior of the induced *constant* voltage associated with the ac Josephson effect is distinctly different from a nonvanishing, time-averaged voltage due to rectification effects associated with asymmetrical current-voltage characteristics. Thus, it is possible to distinguish these two effects by careful and thorough examination of the induced voltage as a function of temperature, rf amplitude, frequency and time. For example, the Josephson-effect-related voltage can be verified visually on the oscilloscope. In a field of a microwave irradiation the direct current (DC) induced due to RJE inside the film sample chaotically oscillates during the change of temperature, amplitude and frequency of a microwave radiation. This is because the common voltage in a line-up of josephson contacts represents a sum of individual voltage each of which oscillates in random manner depending on amplitude and frequency of a microwave radiation.

The main purpose of the given activity consists in the analysis of the results obtained earlier /10/ in thin carbon films, formed by sputtering of spectral pure graphite in an electrical arc discharge (CA-films). Study of the given kind objects has revealed their anomalous electromagnetic properties. The elecromagnetic experiment with carbon foils has been stipulated during the work on carbon strippers /11-13/.

At first, the appearance of switching between low and high resistance has been observed (see Fig.1), which reminds the phase transition between the metal and insulator. However, the metal character of low resistance state (LRS) is doubtful due to a development of RJE in CA-films at room temperatures, which has not been observed earlier in carbon condensates. The given fact allows putting forward a working hypothesis about superconducting character of LRS in CA-films. Indirectly this hypothesis was confirmed by the results of electron spin resonance (ESR) studies of CA-films. In as-produced films ESR signal of conductive electrons clearly was registered and after an annealing - in LRS such ESR signal was absent. There are the other anomalies of electron properties: after transition in a high-resistance condition followed by a jump of resistance by a few orders of magnitude (see Fig.1), the sample without any visible reason (spontaneously) passes in a state with intermediate resistance. Further process is repeated and gains character of oscillations of resistance with the frequency in fractions of Hertz, gradually aiming to pass in LRS a condition.

For more then ten years, passed after publication of activity /10/ the significant efforts on collecting of the facts related to the given problem were undertaken. The papers /14-15/ present the results of studies of anomalous electromagnetics in CA-films by means of magnetic force microscopy (MMF) and DC SQUID magnetization and other methods. This work confirms also the earlier results published in /10/.

To attention of the reader except a main HTS the alternative hypothesis are offered, for example, the analogy with chalcogenide glasses, in which also the effect of resistance switching and peculiar pairing of spins are present (in the modified form).

CA - films, despite of a rather small thickness (*2000A*) represent the object with the complex enough spatial and electron structure. Both two specified structures is described by a set of paradigms. For example, for a spatial structure the role of such paradigms can play both: the theoretical representations and experimental data about a structure and mutual arrangement of various phases of carbon formed at sputtering of graphite in an electrical arc discharge, and also mechanisms of granules formation. As for paradigms of an electron structure, to those it is possible to attribute the theory of a two-dimensional electron spectrum, theory of a BCS superconductivity, and also the theory and experimental data of RJE. Some of paradigms concern simultaneously to both structures. There are the theory and experimental data about granular josephson media, theory of percolation, the effect of irradiation on a structure of CA - films. The gear of comparison of working hypothesizes with the specified paradigms is possible conventionally to call as a filtration, in which the role of filters executed by paradigms. On Fig.2 the scheme of a filtration of representations about processes in CA - films and some corollaries concerning their structure and properties stipulated by used paradigms is shown. The result of passage of each filter is the elimination of some hypothesizes, at the same time those, that could "percolate" gain large reliability in relation to investigated object. Obviously, it is possible to organize double, triple etc. filtration, however every time is necessary to add new filters - paradigms.

The plan of the given article basically corresponds to the scheme of Fig.2.

## 2. Representation about a structure of carbon condensates

By virtue of its polymorphism and numerous varieties for a long time carbon attract the attention of the contributors. Existing nowadays sights on molecular and over-molecular organizations of condensed carbon cannot be considered as completely settled. The structure of various spatially - lacing carbon structural varieties in many respects remains obscure until now, the full clearness is not present even at a level of interatomic *C-C* links in thin carbon films, coals, carbine. In some greater degree of vagueness arise in the relation of higher, large-scale levels of organization of carbon substance. That fact admits only, that there are different forms of carbon: soot, coke, graphite, coals, glass-like carbon, pyrocarbon, carbine, anthracite, diamond, carbon filaments, macroporous adsorbents, composite materials etc.

The singularity of thin carbon films obtained by sputtering is that in them the various varieties of carbons can coexist: diamond with $sp^3$- hybridization of valence electrons, graphite with $sp^2$- hybridization, carbine phase with $sp$ - hybridization. However, all possible types of electron states of carbon atoms cannot be reduced only to three specified. It is possible, for example, $s^2p^2$- hybridization (the model of bivalent carbon) /16,17/, multistage hybridization of already hybrid states etc. The relative stability of tetra-coordinated carbon - singlet biradical /18 /, diamond-like tetra-coordinated amorphous carbon /19/, pentahedral pyramid – semi-sandwich, having a positive charge in cyclopentadienil ring-base of a pyramid /20/ etc. electron states /21,22/ has been find during the computation.

Even the electron structure of graphite is discussed until now /23-27/: it is subjected to a doubt a dynamic symmetry a sigma - bonds in a layer, their planarity, hexagonal symmetry etc. Apart from variety of an electron structure of carbon phases in thin films is observed also the variety of their spatial organization: clusters - granules with a layered graphite internal structure, lamellae of carbine, carbon nanotubes - whiskers, graphite fillets, and diamond-like carbon. Such variety of electron and spatial structures can be exhibited, as will be shown below, in unique electrical, magnetic, optical and mechanical properties of carbon condensates.

The properties of carbon condensates are largely determined by conditions of their deposition. From the vapor phase of ion - molecular beam plasma, generally speaking, the graphite, soot, diamond, carbine can be formed (homogeneously or heterogeneously). Apparently, the type of a formed carbon system is determined nevertheless not thermodynamic but kinetic factors: the intensity of deposited carbon atoms flow, their energy and degree of clustering (cooperation) in a flow, structure and temperature of substrate, pressure and composition of gaseous atmosphere etc. In the activity /28/ it is shown, that at evaporation of graphite by an electron arc the resultant plasma contains particles of various morphology: short circuits of molecules, spherical clusters (fullerenes) and sub-micron particles of graphite (*nc-C*). In the activity /29/ isomers of carbon nano-clusters (*nc-C*) has been investigated by means of the ion mobility spectroscopy method. Computer simulation of approaching geometry of clusters and their isomers in parallel was carried out. It is revealed, that the structures representing a circle from atoms of carbon with a circuit of carbon atoms (tail) attached to it in the best way simulate the found out experimentally isomers. At increase of the size of a cluster the transition from a linear circuit to a circle of carbon atoms happens.

Carbine condensates will be formed in the softest conditions of carbon condensation: at rather high vacuum, small intensity of a flow and low energy of projectile atoms (ions) or their groups, small speed of their deposition. The line-ups grow, apparently, being perpendicular to a substrate, and not sewed together among themselves /30/.

The diamond, on the contrary, is produced in most severe constraints of condensation process - at high energy of projectile atoms (ions). Deeply taking root into the already formed film, projectiles strongly distort its structure, hash atoms of carbon, promoting their chemical seaming. Thus the role of a substrate is insignificant. Chemical active atmosphere and the high pressure also promote seaming of carbon atoms and line-ups with formation of the diamond forms of carbon /31/. It is established also, that at impact pressure up to *120 GPa* the phase transformations of graphite in diamond, and diamond - in metal carbon /32,33/ happen.

In intermediate range, at enough high concentration in a gas phase of atoms with not so high energy the layered polymers of carbon will be formed: pyrocarbon, soot, pyrolitic graphite, with the greater or smaller concentration of interchain and interlayer chemical links (seams) and regularity of their arrangement /30/.

### 3. Relation between electrical resistance and electronic structure of carbon condensates

The irradiation of carbon condensates by high-energy ions give rise to significant changes both the structure and composition of CA-films, and their conductivity.

In the activity of Venkatesan e.a. /34/ the conductivity of polymeric films *PVC* and *HPR-204* was studied under irradiation by ions $Ar^+$ with the energy of *2 MeV* down to a doze $10^{16}$-$10^{17} cm^{-2}$. The resistance of these insulating films decreased up to the value of $\rho_\infty = 3.5 \times 10^{-3}$ *Ohm cm.*

Moreover, in a broad interval of dozes of $10^{14}$-$10^{15}$ $cm^{-2}$ the films are characterized by the temperature dependence of a kind $\rho(T) \sim \rho_\infty exp(\sqrt{T_0/T})$, which is characteristic to charge carry by jumps between conductive islands, separated each from other by the dielectric barriers.

The study of film composition by means of a Rutherford's back scattering method has shown, that almost all carbon remains in film, while other components (in particular, hydrogen) escapes the film. The analysis of a Raman spectrum has shown that the film has a high degree of disorder. At low dozes of irradiation the Raman spectra are lost the sharp singularities and gain the shape with two broad peaks at *1350* and *1580 $cm^{-1}$*, first of which nowadays identifies with so-called nano-crystal carbon (*nc-C*) (i.e. set of carbon clusters with sub-micron sizes), and second – the amorphous carbon /35/. The broad spectrum of the second order was observed also *at 2700-3200 $cm^{-1}$*, which testifies about strong chaotic phase and the absence of *C-H* of bonds at large dozes of irradiation.

The results obtained are confirmed by the later measurements. So, in the activity /36/ ultra thin island of carbon structures were deposited at room temperature on $SiO_2$ - substrate. Then film has been annealed at temperatures *650-700C*. As-deposited clusters of amorphous carbon were characterized by Raman peak at *1500 $cm^{-1}$*. On the other hand the annealed particles of (*nc-C*) exhibited additional peak at *1350 $cm^{-1}$*.

*By the method of Auger electron spectroscopy it was established, that (nc-C) clusters consist of parallel graphite layers with distances between layers ~ 3.4A. It should be pointed that there is no primary orientation of graphite planes in relation to a substrate. Under increasing of the clusters sizes till 20A their coalescence began /36/.*

The pronounced singularities at *1350* and *1580 $cm^{-1}$* also were observed in the graphite under ion irradiation /37,38/. In polycrystalline graphite the ratio of intensities of these two peaks correlates with the mean size of coherent scattering area (CSA) of crystallites. The extrapolation of the data on the case of polymers irradiated with ions in activity /34/ give the size of CSA no more than *30A*. Majority of polymers at high dozes of irradiation has identical Raman spectra. In natural graphite the electroresistance at room temperature is increased with decreasing of the CSA size. The resistance $3 \times 10^{-3}$ *Ohm cm* in natural graphite corresponds to the CSA size ~ *15 μm*.

*Because on the base of Raman spectra it is possible unequivocally to conclude those crystallites with the CSA sizes more then 30A are absent, it is difficult to explain so low resistance of the irradiated polymeric films in terms of graphitization.*

In the activity /36/ with the help of electron energy losses spectrometry (EELS) are found out the nanoclusters (*nc-C*), which sizes grow at an annealing and reach the value of ~ *10A* at the annealing temperature *700C*.

*It is established, that with growth of clusters the general conductivity varies and at the size, greater than 10A the semiconducting conductivity is transformed in the semimetallic one /36/.*

In the activity /39/ the electrical and optical properties of polymer *Kapton H* were studied under irradiation of nitrogen ions with then energy of *1 MeV*. The irradiated area of a film, which in the beginning was transparent, at low dozes of irradiation became black, and at increase of a doze became silvery with a characteristic metal luster. The specific electroresistance of a film irradiated up to a doze $10^{17} cm^{-2}$ is estimated by the value of $10^{-2}$ *Ohm cm*, that is 20 orders of magnitude less than the resistance of nonirradiated film. It is supposed that the changes of electrical and optical properties happen because of decomposition of a film, loss of hydrogen and carbonization of polymer under irradiation. The loss of a main part of hydrogen under irradiation of polymeric film was observed in the activity /40/. A primary ratio of elements in the specified above film was the following *C: Si: H = 9:2:14*. After irradiation with the one-charging ions of argon up to a doze $10^{16}$ $cm^{-2}$ *90 % H, 10%C,* and *5%Si* has been lost. The loss of atoms of hydrogen after irradiation further was confirmed by a technique of measurement based on a nuclear reaction $H(^{15}N, \alpha\gamma)C^{12}$, which has a resonance at *6.385 MeV* with the width of *0.9 keV*. The reaction gives γ-quanta with the energy *4.43 MeV*, in the quantity, proportional to the concentration of hydrogen. The concentration of *H* becomes negligibly small at a doze $2 \times 10^{16}$ $cm^{-2}$. In the activity /41/ the influence of pyrolysis (thermal decomposition) and irradiation by ions of argon with the energy of *2 MeV* on the conductivity of polymeric and carbon films are compared. Both processes result in chemical degradation and reorganization of a structure. On the photographs made with the help of transmission electron microscope (TEM), it is possible to distinguish the formation of the graphite order with the length of a correlation ~ *20A* with the graphite planes oriented in parallel to a surface of a film.

*Conductivity of polymeric films stipulated by thermal decomposition and irradiation by ions tend effectively to the common limit value of $\rho_\infty = 10^{-2}$ Ohm cm /41/.*

The measurement of carrier density at a level of Fermi in the given ultimate state gives rise to the anomalous high value of ~ $10^{23}$ $cm^{-3}$. The crystalline graphite has small overlap of conductive and valence zones *(< 30 meV)* and, hence, the low density of carriers ~ $10^{18}$ $cm^{-3}$. It is supposed, that pyrolytic and irradiated carbon film have the greater overlap of zones (~*1 eV*) at a level of Fermi because of a significant spreading of energy of carriers connected to the short scattering length /41/. In the activity /42/ the relation of electroresistance of irradiated thin hydrocarbon films vs the temperature has been done which is matched with the data of activities /34, 41/:

$$\rho(T) = \rho_\infty \exp(\sqrt{\frac{T_0}{T}}), \quad \rho_\infty = 3.5 \times 10^{-3} \text{ Ohm cm.} \quad (1)$$

The value of $T_0$ depends on a doze of irradiation *D* under the law:

$$T_0 \sim D^{-1.65}. \quad (2)$$

The relation of a type (1) is characteristic for granular media with jump conductivity between the conductive islands.

In the activity /43/ the competition between the effects of a superconductivity and localization of electrons in granular Josephson media are considered. For the samples with full suppression of josephson tunneling between granules the transition of granules in a superconducting state essentially influences the temperature relation and mechanism of jump conductivity. At $T<T_c$ the fulfillment of relation (1), appropriate to influence of Coulomb gap was observed. At $T \ll T_c$ the resistance will increase much more abruptly with the decreasing of temperature and has the relation:

$$\rho(T) \sim \exp(\frac{\Delta}{kT}), \quad (3)$$

where *Δ* –is the value of a superconducting gap.

*The expressions of (1) and (2) considered in the given unit are used further for the analysis of a relationship of electroresistance vs the temperature in CA films.*

## 4. Influence of heating and irradiation on the structure of carbon condensates

The influence of ion bombardment on the structure and lifetime of carbon films has been studied early /11-13, 44-46/.

To shed some light on a structure of films after irradiation the data of activity /47/ can help, in which the two-dimensional ordering of graphite, damaged by ions, was studied. The radiation damage in graphite under implantation of ions can be annealed with the subsequent recrystallization of the irradiated area. Graphitization has two stages: first - at temperatures up to 1500C consists in ordering of atoms in flat layers and formation of so-called turbostrat structure described by absence of natural orientation of layers relatively the hexagonal axes /48/. The second stage at temperatures above 2500C results in formation of three-dimensional ordered graphite structure.

One of the capabilities of structure transformation of hydrocarbon condensates under irradiation is specified in the activity /49/. Here it is shown, that the irradiation by ions with the energies of a few *MeV* calls the breaks of *C-H* bonds with consequent transformation of amorphous structure of carbon and hydrocarbon into graphite one. It is supposed that the destruction of *C-H* bonds and formation of *C-C* graphite bonds is called by secondary electrons arising along a track of ion beam inside the material.

In the activity /50/ the research of micropores in the amorphous hydrocarbon (*a-C:H*) films with the help of pulsed positron beam was conducted. The authors conclude, that in the given films, at least, *3 %* of volumes take the micropores with the diameter about ~ *0.5 nm.* Along a system of such pores the diffusion of molecular hydrogen arising due to destruction of *C-H* bonds during pyrolysis process or ion irradiation is possible.

As it is simple to see from the previous description, the effect of ion irradiation on the hydrocarbon films in terms of their influence on the electrical resistance and their temperature dependence is equivalent to thermal destruction. On the other hand, the result of ion irradiation of polymer films is the complex unregulated structure, in which the effects of a competition of a superconductivity and localization of electrons explicitly considered in the review /43/ are possible. The various techniques of sputtering of carbon condensates with a consequent deposition on a substrate also allow varying over a wide range the structure of the deposited condensate. Therefore it is no wonder, that three specified processes (thermal and radiation destruction and sputtering of carbon with a consequent deposition) create an equivalent microstructure possessing a few universal electromagnetic and optical properties.

*In this sense it is represented interesting to conduct the study of electromagnetic properties of thin carbon films obtained by various methods of a deposition within the framework of representations and the approaches, considered in the activity /43/.*

The similar consideration was carried out for thin carbon films obtained by sputtering of spectral pure graphite in an electron arc (CA-film) /10/. A films with the thickness ~ *2000A*, were deposited on a quartz substrate covered with a soluble sublayer of $C_{18}H_{33}O_2K$. The release of a film from a substrate was conducted by a method of flotation in the distilled water. The as-deposited film have a metal luster and resistance at room temperature of ~ *0.1 Ohm cm*. As the given resistance considerably exceeds the value $\rho_\infty$ (see the expression (1)), the annealing in the vacuum furnace was used at temperature *Ta* ~ 700-1000C within several hours to drop the resistance. In the result of annealing the resistance of a film was decreased down to value $\rho_\infty$ ~ *2-5x10$^{-3}$ Ohm cm*. As it was indicated above, such annealing results in two-dimensional ordering of flat graphite layers /51/, which, as it was established /52/, will form microscopic graphite like crystals with the CSA of about ~ 10-20A chaotically distributed in the amorphous carbon matrix.

The correlation between the microstructure of clusters and their conductivity was considered in the activity /36/. With the help of EELS the dependence of an electron structure on the size of nanocrystal particles of carbon (*nc-C*) was investigated. A structure and distribution on the sizes of (*nc-C*) were determined with the help of Raman scattering, Auger electron spectroscopy and transmission electron spectroscopy.

*With the help of EELS it is established /36 /, that in (nc-C) the transition from semi-metal to a semiconducting state happens at decreasing of cluster size less then ~10A. The given observation can explain observable changes of electroresistance in carbon condensates.*

Thus, CA - film represent the granular media. However, at such rather low temperatures of an annealing still there is no three-dimensional ordering in crystallites along a hexagonal axis, which is observed at *Ta ~ 2500C* /51/. Therefore interplanar distances in crystallites considerably exceed those in natural crystal of graphite, and the decreasing of interplanar distances with the increase of *Ta* reveals a strong correlation with the diffusion removal of impurities from the volume of crystallite /53/.

## 5. Possibilities for HTS phase in CA- films

Proceeding from a justified above thesis about a similarity of structures obtained in the result of thermal destruction, ion irradiation and sputtering of carbon with a consequent deposition, in the unregulated two-dimensional structure of CA films (with a competition of localization and superconductivity) is possible to expect to find out the effects of superconductivity. The similar hope is pertinent also that as it is shown in the review /43/ singularities of conductive properties of granular media are determined by the value of their electroresistance at room temperature. It has appeared, that there is a threshold value of resistance $\rho_c = \pi\hbar/2e^2 = 6.45\ kOhm$, above which the phase coherence does not envelop the whole sample /54/ and the common resistance of a granular line-up does not drop up to zero, thus, the strong superconductivity does not achieved. However at $\rho > \rho_c$ the effects of a weak superconductivity, such as RJE, i.e. detecting of a microwave radiation by a system of josephson contacts–granules are exhibited. In the activity /10/ the conductive properties of thin CA - films were studied. A number of properties is found out which on the first sight seem anomalous, however during closer consideration all these anomalies find their explanation within the framework of representations about granular josephson media which is discussed in the review /43/.

At first, the temperature dependent jump of resistance on ~ *4-5* orders of magnitude was found out (see Fig.1) that gives some idea to assume, that a ground state of a sample was insulator or, at least, the state with very large resistance.

Secondly, the RJE was found out at all temperatures up to ~ *700K* (see. Fig.3). The last circumstance gives the basis to consider a jump of resistance as transition a superconductor-insulator.

*The given conclusion has far going consequences, as it is appeared, that the conductivity of CA films, investigated in an interval of temperatures from 4 up to 700K is stipulated, largely, by Cooper's pairs.*

The CA film can be considered as two-dimensional. This is at first, because of a rather small thickness, and secondly, each granule is formed by stack of two-dimensional graphite layers with significantly larger interlayer distances as compared with natural graphite. Therefore we shall try further to conduct the considerations of such films in two-dimensional approach of an electron spectrum.

The study of electroresistance dependence vs temperature has allowed to establish, that at low temperatures the Mott's law is fulfilled:

$$\ln \rho \sim \sqrt[4]{\frac{T_0}{T}}, \quad (4)$$

with $T_0 = 1520K$, that approximately coincides with a mean Debye temperature of graphite *1600K*. At *T ~ 100K* in the relation there is a fracture (see. Fig.4), which in a similar situation in a system of $La_2CuO_4$ was treated as the indication about change of a system of carriers. Density of states at a Fermi level can be determined similarly to the case of ref./55/:

$$N(E_F) = \frac{16}{a^3 k T_0}, \quad (5)$$

where $a = 10A$ – is the Bohr radius of electron localization.

The expression (5) gives the estimation of $N(E_F) \sim 10^{23}$ (eV)$^{-1}$cm$^{-3}$, that agrees with the data obtained in the activity /41/ if to accept the value of overlap of valence and conductive zones equal to *1 eV*.

In the case of a two-dimensional electron spectrum the Fermi surface is cylindrical and density of states on the one spin can be determined by means of relation /56/:

$$\nu(0) = \frac{z m_{ab}}{2\pi c \eta^2}, \quad (6)$$

where $c$ – is the lattice parameter, $z$ – the quantity of layers in an elementary cell, $m_{ab}$ – the effective mass of a charge carriers. The plasma frequency of electrons will make:

$$\omega_p^2 = \frac{4e^2 z E_F}{c\eta^2} = \frac{4\pi e^2 n}{m_e \varepsilon_\infty}, \quad \eta\omega_p = 3.8\ eV,$$

where $n$ – is the concentration of carriers, $m_e$ – the mass of an electron, $\varepsilon_\infty = \dfrac{m_{ab}}{m_e} = 15$ – the contribution to a dielectric permeability of electrons of internal shells, $E_F = 1$ eV – the Fermi energy.

The value $\eta\omega_p = 3.8$ eV corresponds to $\pi$ - plasmon in a lattice of graphite /57/ which reflects the motion of electrons in graphite layers. In CA - films the impregnation of other phases can be present among them the most frequently can be found the diamond with the plasmon frequencies of 12, 19, 27.7 eV and carbine (16.5; 18.7; 26 eV). As it is possible to see these frequencies are much higher of the value obtained above.

*Thus, it is possible to make a conclusion, that the superconducting phase in CA- films, if those really exists, represents weakly connected graphite layers in granules, located in a matrix made from a carbon material with high resistance.*

The amorphous carbon and such its variety as diamond or carbine can be considered as a matrix material. V.L.Ginzburg's idea of "sandwich" /58/ about increase of a characteristic radius of screening $R_s$ in this connection is represented to be actual:

$$R_s \sim \sqrt{\frac{E_F \varepsilon_\infty}{6\pi e^2 n}}. \quad (8)$$

This happens at the expense of a high dielectric permeability $\varepsilon_\infty \gg 1$ /59/. As it has been shown above $\varepsilon_\infty = 15$ for CA films.

The evaluation of spin density with the help of relation (6) gives the value of $\sim 10^{23}$ cm$^{-3}$ at $E_F = 1$ eV. This is well above the value of $\sim 10^{19}$ cm$^{-3}$, measured by a ESR method in non annealed (as-deposited) CA - films. The given circumstance can testify about difference in carrier systems of annealed (Cooper's pairs of electrons) and non-annealed CA films (jump conductivity).

As it is known /60/, the jump conductivity of localized $\pi$ - electrons in amorphous carbon films can be expresses by the relation:

$$\sigma_{dc} \sim \frac{N(eR)^2 \nu_d}{6kT}, \quad (9)$$

where $N$ – is the number of localized carriers (the number of paramagnetic centers defined by a ESR method), $R \sim N^{-\frac{1}{3}}$ -the length of a jump, $\nu_d$ -the frequency of jumps.

According to the data of ESR measurements $N \sim 10^{19}$ cm$^{-3}$, $\nu_d \sim 10^8$ Hz for non-annealed CA films, then at room temperature we have $\rho_{dc} \sim 10^3$ Ohm cm, that is two orders of magnitude higher the measured value.

*The given disagreement guides us an idea, that an apart from jump conductivity in CA - films other processes of current transfer, for example, superconducting are involved also.*

However this is only indirect indication that will be a subject of furthers checks. As many appearances in CA films can be stipulated by josephson processes, the brief description of josephson media conductivity is adduced below.

**6. Josephson granular media**

Unlike the chaotic metal and dielectric mixes the granular metals have revealed the certain order: they consist of approximately identical granules, separated each from other by the dielectric interlayers. The characteristic scale of non-uniformity is determined by the size of granules. The superconducting coherence between granules can be established by means of josephson coupling. The conductivity and superconductivity of josephson media is determined by quantum tunneling of electrons between granules and depends on the interrelations between the number of characteristic energies /43,61,62/:

1. Difference between energy levels in separate granule:

$$\Delta E \sim (d^3 N(E_F))^{-1}, \quad (10)$$

where $d$ – is the diameter of granule.

2. Energy $\eta \nu_t$ ($\nu_t$ - is the jumps frequency of electrons between granules); the value of $\nu_t^2$ is connected with the probability M of tunneling between granules:

$$M \sim \nu_t^2 \sim \exp(-2\frac{\sqrt{2m_{ab}\varphi}}{\eta}\delta), \quad (11)$$

where $\delta$– is the thickness of dielectric layers between granules, $\varphi$- the effective height of a potential barrier.

3. Energy of charging $E_C$, required for transition of an electron from one neutral granule to another with formation of a couple of opposite charged granules:

$$E_C \sim \frac{e^2}{2C}, \quad (12)$$

where $C$ – is the electrical capacity of granules.

4. Thermal energy $kT$.
5. Energy of josephson coupling between granules:

$$E_j \sim \frac{\pi\eta}{4e^2 R_N}\Delta(T)th\frac{\Delta(T)}{2kT}, \quad (13)$$

where $\Delta(T) = 3k_B\sqrt{T_c(T_c - T)}$ - is the superconducting gap, dependent on the temperature, $R_N$ –the tunnel resistance between granules, which can defined by the expression:

$$R_N \sim \frac{E_a}{e^2 v_t}, \quad (14)$$

where $E_a$ - represents the characteristic energy of misalignment of electronic levels in adjacent granules and can be determined by such phenomena as splitting of levels in small granules, effects of charging and some other reasons /43,61,62/.

The quantum tunneling happens between the states having the identical energy. The misalignment of levels in adjacent granules renders small influence on the conductivity at sufficiently thin dielectric interlayers or at high temperatures, when $\eta v_t$, $kT \geq E_a$. In this case electrons in the granular media moves rather freely by means of non activated tunneling. At rather low temperatures, when $Ej \geq kT$ in such systems the global superconducting phase coherence is established by means of josephson coupling between granules.

The non-uniformity of granular media is exhibited at large thickness of dielectric layers. At the increase of $\delta$ the noncoincidence between levels begins essentially influence the electron jumps at low temperatures. In this case the energy $E_a$ necessary for compensation of misalignment between the levels is gained due to thermal activation. The tunneling becomes activated, and its probability can be determined by the expression:

$$M \sim \exp(-2\frac{\sqrt{2m_{ab}\varphi}}{\eta}\delta)\exp(-\frac{E_a}{kT}). \quad (15)$$

For granulated films the transition to activated conductivity happens at normal resistance $R_N \geq$ *20-30 kOhm* /63,64/. Thus the conductivity of gains has the jump character.

The strengthening of electron localization in granules at the increase of dielectric layer thickness is accompanied by essential changes in superconducting properties of granular metals. First of all, it is necessary to take into account, that in small granules the quantum splitting of levels can result in a smearing of a superconducting gap and suppression of superconductivity. This effect can be neglected in rather large granules, when $\Delta E \ll kT_C$.

In accordance with increase of a thickness $\delta$ of dielectric layers between granules an establishment of a global phase coherence become more and more complicated that result in decreasing of the observable value of $T_C$. At low temperatures in such systems the large influence of quantum phase fluctuations should be revealed, which can destroy a phase coherence /61,65,66/. This result in so-called restore phenomena, i.e. returning from superconducting in a normal state under decreasing of temperature. The given appearance has been observed multiply in granulated /67-72/ and island /63,73-76/ films. It is represented, that the same kind the processes happen and in CA - films /10/.

The return effects are revealed as a minimum of resistance at *T < Tc,* and the minimum resistance usually is not equal to zero. In the activities /72,74,77/ the inspected change of josephson coupling between granules has been conducted. These enables to monitor the evolution of return effects at transition from homogeneous metal to thin films consisting of metal islands separated by dielectric. Thus the electrical contact between separate islands is not guaranteed, and the conductivity of such films is determined by a competition between jump and percolation mechanisms. In the activity /74/ such transitions from metal to a dielectric system was achieved by inspected change of a mean thickness of ultra thin films, and in /72/ the similar effect was reached by means of change the constant voltage applied to a granular film of constant.

The theoretical understanding of return appearances is undertaken in the activity /61/. According to /61/ the returning mechanism it is possible to understand qualitatively, using a ratio of indeterminacy between the number of quasiparticles *N* in granule and phase *φ*:

$$\Delta N \Delta \varphi \geq 1. \qquad (16)$$

1. In the case of $E_C \sim kT_C$, at $T \gg Tc$ the granulated film has not revealed the activated conductivity. At *T* decreasing for all adjacent granules can begin to be satisfied the condition:

$$E_j > E_c + kT. \qquad (17)$$

   Therefore inside the system the global phase coherence of electrons is established. At the further decreasing of *T* the condition $E_C \gg kT$ will begin to be satisfied, which means the transition to activated tunneling of one-particle excitations. According to (15) the probability of such tunneling is exponential decreases with decreasing of temperature, that results in decreasing of *ΔN* in (16) and therefore *Δφ* increases. This means the violation of phase coherence of electrons in various granules. The violation of phase coherence of electrons in adjacent granules under the conditions of significant decreasing of quantity of uncoupled electrons in superconducting granules in the conditions of temperature decreasing means the transition of a conglomerate of granules in a dielectric state. I.e. in the case of *Ej ~ Ec* at high *T> Tc* the system behaves as usual metal. Then at decreasing of *T* the system passes in a superconducting state, but at T << Tc it becomes insulator.

2. The electron processes in CA - films /10/ can be explained by full suppression of coupling between granules (*Ec >> Ej*), when character of activated conductivity essentially varies due to appearance of superconducting gap $\Delta(T)$ in a spectrum of electrons /61/. This is the dielectric gap. According /61/ in such conditions two mechanisms of activated conductivity act: the activated tunneling of Cooper's pairs and one-particle excitations with energies of activation *Ec* and $\Delta(T)$ correspondingly. The main contribution to conductivity should give the process with smaller activation energy.

   At *Ec* < $\Delta(T)$ the main contribution should give the jumps of Cooper's pairs which at *T << Tc* determine the temperature dependence of electrical resistance:

$$R(T) \sim \exp(\frac{E_c}{kT}), \qquad (18)$$

   which, apparently, also is observed in CA films.

It is interesting to note, that the condition *Ec >> Ej* and $R_N > \dfrac{\eta}{4e^2}$ means, that the charge of granules is a "good" quantum number /78/, when *2e* periodicity of an electric current in a circuit of granules as functions of a granule charge is exhibited.

It is represented, that the experiment on check *2e* - current periodicity could spill of light on a problem of existence room temperature superconducting state in CA - films investigated in /10, 14-15/.

On a curve of *R(T)* (see. Fig.3) a bending in the region of *T = 100K* can be seen. The high-temperature part curve behind a point of inflection, probably, is stipulated by motion of quasiparticles with the activation energy of $\Delta(T) \sim 0.03\ eV$. The another part of curve at temperatures lower than *100K*, apparently, is described by the expression (18) and is stipulated by the motion of Cooper's pairs. The evaluation of *Ec* from the slope of the curve gives the evaluation of $E_C = 0.00324\ eV$. Then with the help of expressions (12) the diameter of CA film granules *d ~ 6 microns* is obtained, that will be agreed with the data of ESR measurements indicating the existence a net of josephson contacts with the cell size being in an interval of *1 - 10 microns.*

The result obtained gives us a reference point in studies of structural inhomogeneities and phase structure of CA films, which can explain their exotic electronic properties. Probably, non-uniformity at a level of *~ 10 microns* arise due to a net of microcrystalline embedments of carbine /79/, which grow with increase of annealing temperature. As it is shown in the activity /80/ carbine occur as thin layers-lamellae with the sizes of *~ 3-15 microns* interleaved with graphite.

The evaluation of $E_C$ value with the help of relation (14) allows to determine the tunnel resistance between granules, which gives the bottom evaluation of resistance in a normal state $R_N \geq 10^8\ Ohm$. Whence at the $T_C = 350K$ from a relation (13) at once the value of josephson coupling energy can be obtained $Ej \sim 3\times10^{-5}\ eV$. However, if to assume, that the non-uniformity at a level of *~ 10 microns* originates due to spontaneous impregnation of carbine lamellae, the problem arises with the interpretation of small-scale net of josephson contacts with the sizes about *~ 10A*, which is detected with the help of microwave - radiation /10, 14-15/. It is possible, that the given net arises due to josephson interaction of graphite layers in granule just as it is circumscribed in the review /81/. Other conceivable capability of small-scale net origin is the formation of twin defects in crystal structure. In the activity /82/ it is shown, that misalignment of crystallographic axes of adjacent granules on a few degrees ($\geq 5^0$) can gives rise to the formation of intergranular josephson junction. However, at such explanation is not clear, why josephson interaction of adjacent granules is so weak. Most likely, the structure of CA films represents a conglomerate from josephson granules, separated by high-resistance dielectric barriers, which can play the carbine (its specific resistance is high enough $\sim 10^6\ Ohm\ cm$) or another insulators such as diamond-like or amorphous carbon. In this case the sizes of insulator interlayers are determined by the sizes of granules *(~ 10A)*.

*The origin of a large-scale grid of josephson contacts with the sell size ~ 10 microns can be determined by percolation character of a current in a small-scale network of granules /83/.*

**7. Possaible analogy with the chalcogenide glasses**

Some property of CA films very much remind those of glass carbons. So the electrical resistance of glass carbon changes in an interval *0.001 - 0.02 Ohm cm*, and the density varies in the range of *1.3-1.8 g/cm$^3$* /84/.

The electron properties of glasses generally are very various. So, for example, in chalcogenide glasses the appearance of switching /85/ is observed. Besides it was established, that chalcogenide glass are not sensitive to impurity and do not contain uncoupled spins /86/. The similar property, as it is visible from the previous description, is inherent also for CA - films /10, 14-15/. The impression from this analogy amplifies because of presence of switching effect in glass carbon /87/ and carbon deposits /88/. It is turning out that electroresistance of these systems at some threshold voltage (2-10 V) spasmodically drops approximately on an order of magnitude. As the conditions of obtaining of carbon films in /88/ are rather close to those /10/, it is possible to assume a similarity of switching effects in these two cases. However there are also essential differences. At first, the jump of resistance in /10, 14-15/ makes 4-5 orders of its magnitude (see Fig.1). Secondly, the jump of resistance in /10/ in difference from /88/ regularly depends on temperature.

Nowadays it is considered to be well established, that the effect of switching in chalcogenide glasses has an electron nature /89/. Probably, the most exotic property of chalcogenide glasses is the full absence of spin density. Even if each atom will form locally all required valence bonds, it would be possible to expect that at room temperatures thermal energy cause the rapture of some from the most high-energy bonds, as it happens in the majority of other substances. In

1975 F.U. Anderson has noticed, that if two electrons with opposite spins localized near to atom of chalcogenide glass, are effectively attracted each other, than the ground state of the system will have a zero density of uncoupled spins. This effect, as it is known, lies in the basis of superconductivity.

Further it was established, that the localized state in chalcogenide glasses occur not due to of substance disorder, but because of quite certain defects, which are the pairs of atoms with alternated valence (see. Fig.5). It was shown /90/, that it is possible to explain a broad spectrum of properties of chalcogenide glasses if to assume the existence of effective attraction between localized electrons. To understand the properties of chalcogenide glasses, it is enough to realize that in such systems a unique electroneutral defect exists which is the selenium atom with the extra bond (three-valent atom), instead of free valence (univalent atom). The presence of the last type of defects would result in occurrence in chalcogenide glasses of uncoupled spins because the removal of an electron from one such defect and the pairing with an electron of other defect results in essential increase of energy at the expense of an arising electrostatic repulsion of electrons. The removal of an electron from the three valent selenium atom results in a state with low energy, since in this case selenium atom becomes structural similar to atom of V group in terms of its electron configuration. However, if other three-valent atom of selenium catches the remote electron, the condition with high energy (two electrons on - neighborhood) is formed. The condition with low energy can be formed after catching of this electron by one of nearest neighbors of three-valent atom with a subsequent rapture the bond between them. As a result the selenium atom which had received an electron remains connected only with one nearest neighbor (see. Fig.5a). In this condition the selenium atom is electronically similar to the halogen one from VII group with all its intrinsic bonds.

So, the transition of an electron from one trigonal-connected center to other with rapture of bond on the last center decreases the energy of substance. This is the reason of effective attraction between localized electrons. It is easy also to see, that each such produced alternated positively and negatively charged centers has only coupled spins.

The explained above concept of active centers in chalcogenide glasses allows, on the one hand explaining the mechanism of switching. Actually, the conductive state is reached only then, when all present in glass positively and negatively charged traps are filled with carriers exited by an applied electrical field. The filling of traps results in sharp increase of lifetime of the injected carriers. If earlier it was much less time needed the carriers to cross the whole thickness of a film, after filling of traps it becomes much greater of this time. This gives rise to the sharp increase of a current at smaller voltage, i.e. the appearance of a conductive state.

On the other hand, the given approach allows to give the explanation of the granular structure mechanism formation inside the chalcogenide glasses. The stability of a granular structure is stipulated by low surface energy of clusters, which prevents their coalescence. As it was shown above, at presence of both two-valent and tree-valent atoms of selenium the process of formation of active centers with alternated valency becomes energetically expedient. In the activity /91/ on the base of topological reasons it is shown, as to what manner a such processes result in formation of a granular structure. At a certain ratio the component of an alloy a condition with a high saturation of bonds in clusters and absence of bonds on their boundaries are realized. The similar structures experimentally were registered in the activity /91/.

Despite of all logic orderliness of the theory of active centers and similarity of many electron processes in chalcogenide glasses to that in CA films the direct extension of conclusions of the given theory on CA- films is represented premature for several reasons.

At first in CA films effect of detecting of a microwave radiation with the help of RJE is observed /10, 14-15/, that must be connected with a superconductivity.

*On the contrary, it would be possible to try to find the RJE in chalcogenide glasses and glassy carbon.*

Secondly, D.Adler /86/ describes the argument illustrating on his sight, why tetrahedral amorphous substances, such as carbon, silicon and germanium cannot create topological structures, similar to them, that we have observed in chalcogenide glasses. On his opinion only by means of s-and p- electrons of elements of IV group it is impossible to form five bonds. However last reason can appear inconsistent in the case of carbon. This is because it is impossible to reduce all possible types of an electron states of atoms of condensed carbon only to three kinds stipulated by *sp -, sp$^2$ -, sp$^3$* - hybridization. The computational way shows relative stability of a pentahedral pyramid (the state with the five bonds!) – half-sandwich, with a positive charge /17/.

*Therefore concept of active centers nevertheless can be considered as alternative for superconductivity at explanation of anomalies of electron properties of CA films.*

## 8. Applications of ESR method for studies of CA films

In the light of absence of paramagnetic centers in CA films the title of the given paragraph seems to be error. Really, despite of repeated attempts it was not possible to find out in the annealed CA films the free radicals, uncoupled spins, conductive electrons which are confidently detected in graphite and some other carbon materials /93/. The reached limit of sensitivity on uncoupled spins at room temperature has made < $10^{11}$ spins / samples. Such impurities as *Cu, Fe, Mn*, and also the rapture bonds were not registered /10/. However the ESR method nevertheless has been used at studies of HTS ceramics. So in the activity of V.V.Kveder etc. /94/ the singularities of a microwave - losses inside HTS ceramics in a magnetic field in the range ~ *9 GHz* were studied. The sample with the volume of ~ 1 mm$^3$ was placed in a maximum of a microwave - field inside a resonator of ESR spectrometer, and the microwave power *R* absorbed by a sample and its derivative on the magnetic field $dR/dH_0$ was measured. In the superconducting region the noise-like change of $dR/dH_0$ was observed on the background of monotone change of $R(H_0)$. Thus the good recurrence has been observed at repeated scanning of a field. The interpretation of the given appearance /94/ is that the researched ceramics represents a network of a superconducting phase, which cells contain the pores or non-superconducting inclusions. The contacts of superconducting grains can have properties of weak links, i.e. the quanta of a magnetic flow can slip through such links in the superconducting cells without pinning. So there is a "SQUID's" network and the observable relations of $R(H_0)$ correspond to a quantum penetration of a flow in such network. The Fourier image of $R(H_0)$ has a number of peaks appropriate to the areas of cells:

$$S = \frac{\Phi_0}{\Delta H_0} \ . \qquad (19)$$

In the last expression $\Phi_0$ – is the quantum of a magnetic flow, $\Delta H_0$ – the width of peak.

Similar kind the experiments were carried out with CA films. The appropriate relation $dR/dH_0$ is shown on Fig.6. The analysis of Fig.6 data shows, that the characteristic size of a josephson cell lies in the range of *1-10 µm*. This is well above the graphite crystallites sizes of *10-100 A* which can be grown inside this films during the annealing time of ~ *10 hours* at the temperature of ~ *1000C.*

From a comparison of ESR spectra of as deposited and annealed at *1000C* film the increase of amplitude of $dR/dH_0$ oscillations is revealed. As it is known /30/, at an annealing the sizes of carbine clusters in carbon condensates grow. It is possible to assume that there is a correlation between these two processes.

*Besides it was revealed that at decreasing of a microwave power the amplitude of $dR/dH_0$ grows and the oscillations become clearer. So the increase of microwave - signal power suppresses the oscillations. This conclusion is represented to be strange without the assumption that oscillations – is the property internally intrinsic to a SQUID network in CA - films.*

One of the interesting capabilities of noise-like spectrum application is circumscribed in the activity /95/, where it is shown, that the noise-like signal disappears at the magnetic fields $H > H_{C2}$ at appropriate temperature (see. Fig.7). On the base of given effect the way of precise determination of $H_{C2}(T)$ in ESR experiment is offered.

As to CA - films, the attempt was undertaken to conduct similar experiment in Institute of Chemical Physics of RAS. The probable changes of noise-like ESR - signal were investigated at scanning in a magnetic field in an interval *0.1-5 Tesla* at temperatures of *100* and *350K.* Unfortunately the noise-like signal changes (vanishing) has not been observed. The probable reason consists, apparently, that in quasi two-dimensional case $H_{C2}$ is very high (up to ~ *1000*

*Tesla* at *T = 350K*). Besides as it is shown in the activity of Busdin and Bulaevsky /96/ $H_{C2}$ can have the non-zero value even at *T ~ 2T_c*.

*Therefore under planning the tasks on the future, it is necessary to provide a capability of a prolongation of ESR experiments with the noise-like signal at temperatures up to 700C.*

### 9. Quasi two-dimensional superconductivity and the model of Josephson interaction of the layers

The attempt of two-dimensional representation of an electron spectrum in CA films has been presented above. Moreover, the basic computational parameters were determined just in this representation. Two-dimensionality of an electron spectrum can be stipulated by the structure of films composed by a set of parallel graphite layers, which may be integrated in domains or granules, divided by dielectric interlayers. The distances between layers in domains considerably exceed those in crystals of graphite, hence, it is possible to expect weak interaction between the next layers, that can result in josephson interaction of layers /97/. The applicability of the given approach can be justified as follows /98/. Let's consider the effect characterized by the energy $\varepsilon_0$ on one electron. Then in relation to this effect the system of electrons can be considered to be a two-dimensional, if the electron energy $\varepsilon_\perp$ connected with its motion between layers, is much less, than $\varepsilon_0$. For a superconducting condition characteristic energy $\varepsilon_0$ is the value of a superconducting gap of $\Delta(T)$. At jump mechanism of conductivity across the layers

$$\varepsilon_\perp = \frac{\eta}{\tau_\perp}, \quad (20)$$

where $\tau_\perp$ - is the time of electron jump between two next layers. The condition of quasi two-dimensional superconductivity can be expressed as follows /99,100/:

$$\frac{\eta}{\tau_\perp} \ll \frac{\Delta^2(T)}{T_c} \cong T_c - T . \quad (21)$$

If the condition (21) is fulfilled than consideration of those superconducting properties, which are not connected immediately to motion of electrons between layers, can be made in two-dimensional approach. At the same time, the motion of superconducting electrons between layers can be considered as josephson motion. Really, under condition of (21) interlayer currents cannot destroy a superconductivity inside the layers, and the current density $j_{n,n+1}$ between the layers *n* and *n + 1* can be calculated under the theory of perturbation. In a purely two-dimensional system the phase fluctuation can destroy the long-range superconducting order. However transitions of electrons between layers completely suppress such fluctuations, and the order parameter inside a layer practically does not depend on probability of transitions of electrons between layers /98/.

Like the case of usual josephson junctions, the current $j_{n,n+1}$ in this situation is determined by the formula:

$$j_{n,n+1} = j_c (\Delta) \, Sin \, (\varphi_n - \varphi_{n+1}), \quad (22)$$

where $\varphi_n$ – is the phase of order parameter in a layer *n*. Let's mark, that for the occurrence of josephson effects in usual tunnel junctions it is required, that the probability of passage of an electron through a barrier in a junction should be much less than unity /101/. In layered compounds the occurrence of josephson interaction of layers is the consequence of a stronger condition (21). This is because that in usual josephson junctions the superconductivity is more stronger volumetric effect than the surface effect of tunneling of electrons between superconductors. In layered systems a thickness of layers is essentially atomic and both effects (superconductivity and tunneling) have the identical dimensionality.

The fulfilling of a condition (21) in CA films is possible to estimate as follows. Let $d \sim 4A$ – is the interlayer distance then $\tau_\perp$ is connected to a diffusivity $D_\perp$ and conductivity $\sigma_\perp$ by the relation:

$$\sigma_\perp = e^2 D_\perp N(0), \qquad (23)$$

$$D_\perp = d^2 \tau_\perp^{-1}. \qquad (24)$$

As $\tau_\perp^{-1}$ it is possible to try to take the frequency of carriers jumps between conductive "islands", which has been determined by means of ESR method for non annealed CA films $\tau_\perp^{-1} \sim 10^8$ Hz. As $\sigma_\perp$ it is possible to take the value of $\sigma_{dc} = 0.001$ (Ohm cm)$^{-1}$ for the jump conductivity obtained with the help of the expression (9). Then the values of diffusivities determined from the equations (23) and (24) will be rather close each other $D_{\perp 23} \sim 6 \times 10^{-8}$ cm$^2$/sec and $D_{\perp 24} \sim 16 \times 10^{-8}$ cm$^2$/sec correspondingly. And if we make $d = 3A$ they will be practically identical. It is represented that the comparison carried out gives us one more argument for the benefit of josephson interaction of layers. As for the relation (21), as it is easy to check this condition is executed practically everywhere below than $T_C$, since $\dfrac{\eta}{\tau_\perp} = 7 \times 10^{-4}$ eV, $T_C = 0.03$ eV.

In conditions, when the external fields cannot change the value of order parameter module in a layer (electrical field is perpendicular to layers, magnetic field is parallel to them), in the quasi two-dimensional superconductors the josephson-like effects can be observed. Essentially the layered compounds under condition of (21) represent a set of josephson junctions situated in parallel with the superconductors of atomic thickness. Therefore at a constant potential difference $U$ applied perpendicular to the layers, in a circuit can be observed the josephson alternative current with the frequency $\nu = 2eU/(\eta N)$, where $N$ – is the number of layers, between which voltage $U$ is applied /102/. In quasi two-dimensional superconductors there should be also to exist their own oscillations appropriate to josephson plasma oscillations /97/.

The constant voltage created on josephson junction under microwave irradiation with amplitude $V$ and frequency $\nu$ can be expressed by the relation /103/:

$$V_{dc} = RIJ_n(\dfrac{2eV}{\eta \nu}) Sin\Phi_n, \qquad (25)$$

where $R$ – is the resistance of a junction, $I$- the current in a circuit containing a junction.

As it is visible from the above-mentioned expression $V_{dc}$ should be a periodic function of $V$ and $\nu$. The periodic dependence on $\nu$ in CA films was observed, at the same time $V_{dc}$ depends on $V$ only linearly. Such discordance can take place, if the argument of a cylindrical functions $J_n$ is small, that can be explained by a smallness of amplitude of alternative voltage $V$, induced by microwave - radiation in a separate layer.

The ways of strengthening of josephson interaction in CA films are vary depending on mechanism of this interaction. In the case of the granular mechanism it is necessary to undertake the efforts to increase the conductivity of high-resistance phase. This can be reached by means of films annealing. As it is known from experience, the conductivity of films grows in the beginning of annealing, and then leaves on a constant plateau or even begins a little to decrease with increase of duration of an annealing. In the case of josephson interaction of layers the conductivity of films can be increased by means of intercalation.

## 10. Magnetic properties of CA films

For quasi two-dimensional superconductors, when $r=\eta/\tau_{\perp}T_c \ll 1$ (note, that $r \sim 0.02$ for CA films) the sharp growth of critical field $H_{C2}$ in parallel to the layers is characteristic, at temperature approaching the point $T^* = (1-2)T_C$. In the vicinity of $T^*$ the value of $H_{C2}$ for a superconductivity along layers is determined only by paramagnetic limit, and at $T \rightarrow 0$ the relation for $H_{C2} \sim 15T_C$ can be obtained, where $H_{C2}$ is expressed in kilooersted /96/. The appropriate value for CA films at the $T_C = 350K$ gives $H_{C2} = 540$ Tesla. This value is huge and it is seems unreal. Therefore we shall try to estimate $H_{C2}$ from the theory of granular media. The coherence length can be expressed as follows:

$$\xi = \sqrt{\frac{\pi D_{II} \eta}{8kT_c \ln \frac{T_c}{T}}}, \qquad (26)$$

where as the $D_{II}$ it is necessary to use a diffusivity of carriers in a high conductivity state. For CA films $D_{II} \sim 5 \times 10^{-3}$ cm$^2$/sec and at $T = 300K$ we can obtain $\xi \sim 5A$. From a theory of granular media point of view at $\xi < d$ the josephson interaction between granules is so weak, that granules can be considered as the independent each from other zero-dimensional superconductors /104/. Then the upper critical magnetic field will make:

$$H_{c2}(T) = \frac{\Phi_0}{2\pi\xi^2}, \qquad (27)$$

where $\Phi_0 = 2 \times 10^{-7}$ Gs cm$^2$ – is the magnetic flux quantum. For the case of CA films the value of $H_{C2}$ (300K) ~ 1000 Tesla can be obtained. Thus, both theory of granular media and theory of josephson interaction of layers give the huge value of the upper critical magnetic field.

## 11. Fluctuation conductivity

In the granular media of zero-dimensional superconductors the effects of superconducting fluctuation should be strong. In the activity of Decsher etc. /104/ it is shown, that the fluctuation superconductivity of such systems is determined by the expression:

$$\frac{\Delta\sigma_{fl}}{\sigma_n} = \left(\frac{\varepsilon_{co}}{\ln\frac{T}{T_c}}\right)^2, \qquad (28)$$

where

$$\varepsilon_{co} = \left(\frac{d_o}{d}\right)^{\frac{3}{2}}, \quad d_o = \left(N(E_F)kT_c\right)^{-\frac{1}{3}}. \qquad (29)$$

For CA films $d_o$ makes ~ 7A. It is not difficult to see, that in the case of CA films near to the $T_C$ the fluctuation conductivity can give the value of $\frac{\Delta\sigma_{fl}}{\sigma_n}$ ~ 700. Therefore the research of fluctuation effects is a serious problem of the further research of electron anomalies in CA films.

## 12. Capabilities for increase of conductivity at implantation and intercalation

All above mentioned, and also the accumulated experience /30,52,105/ give rise to a conclusion, that the structure of CA films represents a set of weakly interacting two-dimensional graphite-like layers. The intrigue of the previous discussion was twisted round a structure of inhomogeneities of CA films, which cause the RJE. Whether such inhomogeneities are the granules with the characteristic size ~ *10A* or josephson interaction of layers has take place?

*Actually the answer this question determines the further script of increase of normal conductivity and therefore the prospect to obtain the volumetric (strong) superconductivity. If the structure of CA films is granular, it is necessary to increase the interaction between granules by means of implantation into the insulation interlayers the impurities, which can reduce the resistance. In the case of josephson interaction of layers, the increase of normal conductivity can be achieved with the help of intercalation.*

Due to their anisotropy the graphite crystals can produce a research capability of the most interesting effects of physics of a solid state in their two-dimensional appearance. It is wonderful, that the anisotropy of graphite compounds can be in a strongest degree increased under intercalation, - i.e. introduction of atoms or molecules between graphite layers /106/.

*As the graphite-like CA films revealed the josephson effects, we obtain a unique capability of their research as a system with two-dimensional motion of carriers.*

The intercalation for a long time gives some hope to realize the exciton mechanism of HTS which has been proposed by Little /5/ and Ginzburg /107,108/. The class of intercalate compounds practically is not limited, since there is a large number of molecules and atoms, which can be entered into layered graphite-like crystals /109-111/. In particular, the crystals with alternated metal and semiconducting layers are, apparently, most perspective in view of a realization of exciton superconductivity. Within the framework of the given approach it is necessary to create a system, in which the conductive layers should be connected to molecular or semiconducting layers. The polarization of molecules or semiconductor by conductive electrons should result in the effective attraction of conductive electrons. As the value of the $T_C$ is proportional to the excitation frequency of a polarized system, at the replacement of phonon (Debye's frequencies) with the exciton (frequency of the order *0.1-0.3 eV*) mechanism the pronounce increase of the $T_C$ could be expected. Thus the two-dimensional systems have the advantage that the fluctuations destroying the long-range superconducting order in such systems are less essential as compared with the one-dimensional one/112/.

Weak Van der Waals bindings between the layers in graphite-like compounds and, in particular, in CA films allow to enter into space between these layers some atoms or molecules. To the present time the large number of graphite compounds intercalating with atoms and molecules is obtained, which not full list is adduced in a Table 1 /109/.

The application of this intercalates to CA film limits by strong interaction of majority from them with an atmospheric air, decomposition at the elevated temperatures, and also carbonization of intercalates. In the case of the donor intercalates the conductivity is increased, both in layers, and between layers, in the case of acceptors intercalates the conductivity between layers decreases /109/.

The brief digression in a history of research of carbon composite materials shows, that the intercalation is rather perspective way to increase the conductivity and even the superconductivity. Really, the fullerenes become conductors /113/ and superconductors /114/ at doping by atoms of alkaline metals with an optimum structure $K_3C_{60}$. The graphites, as it is known, have two stages of intercalation with potassium /115/: first stage with an optimum structure $KC_8$ and the second stage - $KC_{24}$. The recent researches have shown, that intercalation of composite crystal made from carbon nanotubes gives rise to the metal conductivity in an optimum composition of $KC_{16}$ /116/.

## 13. Possible applications of CA films

The high sensitivity of tunnel superconducting junctions to external electromagnetic radiation due to existence of RJE especially is brightly exhibited in Josephson media of CA films due to possible synchronization of great many of junctions. The effect of synchronization was observed in large number of the experiments which have been carried out with so-called BPB -ceramic ($BaPb_{1-x} Bi_x O_3$) /117-119/. In particular, in film samples the occurrence of both the voltages $U \sim 2\ mV$ and a hysteresis loop on voltage-current characteristics (VCC) under microwave irradiation with frequency *1.8 GHz* was observed /114/. Besides the VCC with discrete values of a current $\Delta I$ were observed /120/, and at resistive load $R > 1 kOhm$ the voltage jump $\Delta V = R \Delta I$ corresponds to the value of superconducting gap $2\Delta (T) = 2.8 meV$.

*The results of experiments described above and likeness of processes in BPB and CA structures can be the indication about prospective use of CA films as a material for elements of highly sensitive receivers of electromagnetic radiation, microwave generators and various kind of switches. And in the case of CA films such devices do not require the cryogenic cooling, as the detecting of microwave is occurred at room temperature /10,14-15/. Some of these properties of CA films uses in projects of real noncryogenic devices /121-123/.*

The important characteristic of microwave detector sensitivity is so-called volt - watt relation (VWR) i.e. the ratio of detecting constant voltage to the power of microwave signal. In the activity /124/ the detecting characteristics of thin-film bridges made from ceramics *YBaCuO* under microwave irradiation with a wavelength ~ of *8 mm* are circumscribed. At nitrogen cooling VWR for these films is about $5 \times 10^5\ V/W$. For CA films the given value makes of ~ $10^6\ V/W$ at room temperature, that does not concede to sensitivity of the best cryogenic superconducting detectors.

In the ref. /125/ on BPB films at passing of mutual orthogonal transport currents $I_x$, $I_y$ the increase of $I_x$ above critical values resulted in shift VCC $V_x - V_y$ on the value of $\Delta V$ which is multiple to $2\Delta (T)/e$. The given structure allows realizing the power source with zero impedance, which can be used as a new logic element of so called "*Josephson computers*".

One more interesting application of CA films can be non-cryogenic Josephson detector of gamma - radiation for registration of a neutrino and dark matter /126-127/. In their recent work /128-129/ Christian Beck and Michael C. Mackey proposed to extract the density of dark energy in the universe from the measured spectra density of current noise in Josephson junction array. In agreement with today's theory the dark energy density $R_{dark}$ of the order $4\ GeV/m^3$ gives rise to the upper cutoff frequency in the noise spectra of JJA of $1.7\ 10^{12}\ Hz$, which increases proportionally with the critical temperature of JJA structure. So it is believed the CA- film JJA structure with the possible critical temperature in the range *700-1000 K* /10,14-15/ could substantionally increases the upper cutoff frequency of JJA and consequently widen the searching range for the $R_{dark}$.

## 14. Summary

To summarize all described above, it is possible to tell, that the basic attention in the given activity was given to discussing of anomalous electromagnetic properties of thin carbon films obtained by sputtering of spectral pure graphite by means of electron arc discharge (CA films). During the research it was established, that in CA films the appearance of switching between low and high electrical resistance takes place. Such effect reminds the metal-dielectric transition. However the metal character of low-resistance state is represented doubtful due to detecting of microwave radiation in CA films stipulated by RJE even at room temperatures. The given circumstances have allowed putting forward a working hypothesis about superconducting character of low-resistance state of CA films.

The main content of the activity is devoted to the substantation of the given hypothesis, therefore the model of granular media was considered explicitly. It is shown, that the given model describes satisfactory the whole spectrum of anomalous electromagnetic properties of CA films.

Other acceptable explanation represents the model of Josephson interaction of layers. The given model saves all advantages of granular media and also gives the explanation to the appearance of inhomogeneity at a level of the sizes in some angstrom due to characteristic layered structure of graphite crystallites in CA films. The difference between a layered structure of graphite and conductive phase in CA films thus can be explained by large interlayer distances in CA films as compared with the crystals of graphite. Besides due to random orientation of graphite particles, deposited by arc discharge the peculiar granular structure will be formed, chosen by orientation of layers in each deposited granule. Really as it is pointed in the activity /79/ disorientation in units of degrees between the layers of neighbor granules already creates the Josephson tunnel junction. Obviously two competing models are jointed in the given point.

Besides some alternative suppositions about the conductivity of CA films, such as the analogy with chalcogenide and carbon glasses and also the hypothesis of fluctuation superconductivity were considered also.

Summarizing the results of the given activity, it is possible to tell that despite of some progress in understanding of anomalous properties of CA films, nevertheless remain many problems far from their solution. The basic problem nevertheless there is an ambiguity of existence of a superconducting phase in CA films. Therefore it is very important to continue searches of experimental capabilities for elimination of ambiguities in explanation of the results obtained. With this purpose under planning of future research, it is possible to offer the following experiments, which, it is believed, has been explained unequivocally:

- Searches of Bloch oscillations of a current as functions of a granule charge with period *2e*
- ESR experiments at the elevated temperatures up to ~ *700C* with the purpose to determine the upper critical magnetic field $H_{C2}$.
- Measurements of RJE at high temperatures ~ *700C* and high intensities of a microwave signal.
- Measurements of electroresistance of CA films at high temperatures
- Measurements of electroresistance at various applied voltage
- Precision measurements of a magnetic susceptibility of CA films with the help of SQUID magnetometer
- Check of the Wideman-Frants law fulfillment
- Measurements of an infrared absorption in CA films
- Experiments on measurement of CA films electroresistance at extreme small electrical currents ~ $10^{-9}$ *A*
- Measurements of dependence of switching voltage vs distances between contacts
- Experiments on intercalation of CA films for strengthening of interaction between layers
- Researches of a cluster structure of CA films by a method of scattering of cold neutrons
- Study of an elementary composition of CA films by means of a method of laser mass spectrometry

In summary I express sincere thanks to A.I.Golovashkin, taking the trouble to read the manuscript of the article and to make a number of the useful remarks. Also I'd like to express many thanks to my wife G.S.Lebedeva for the help in producing of the figures.

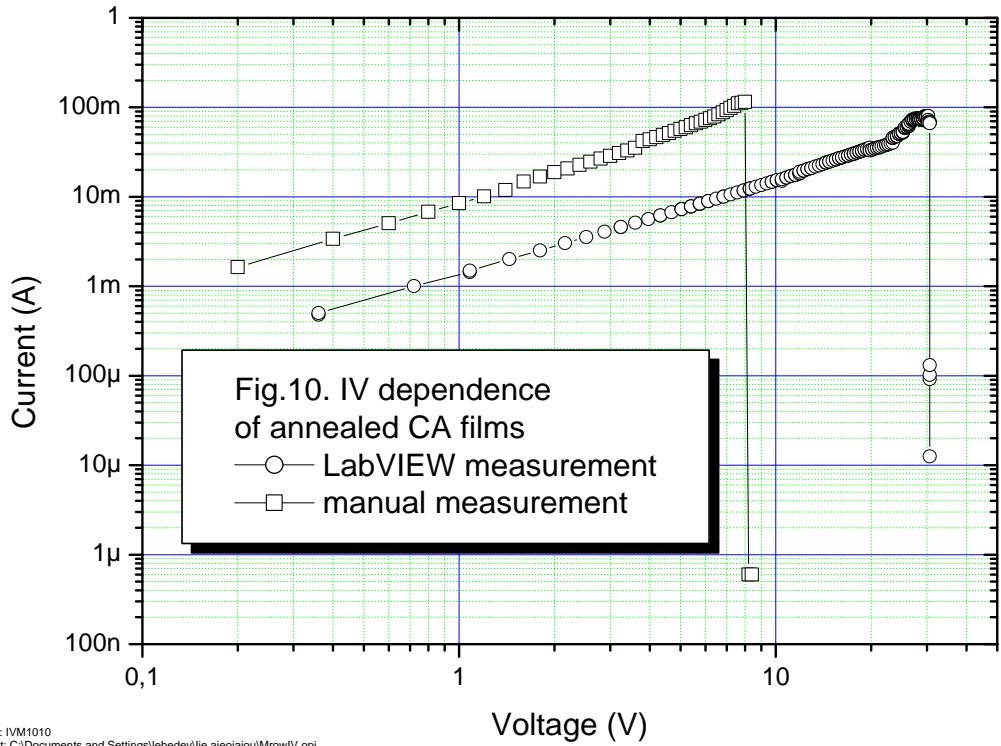

Fig.1.　　Current-voltage characteristics of thin CA- films.

Fig.2 Block diagram of representation filtration about the processes inside the CA films and their consequences.

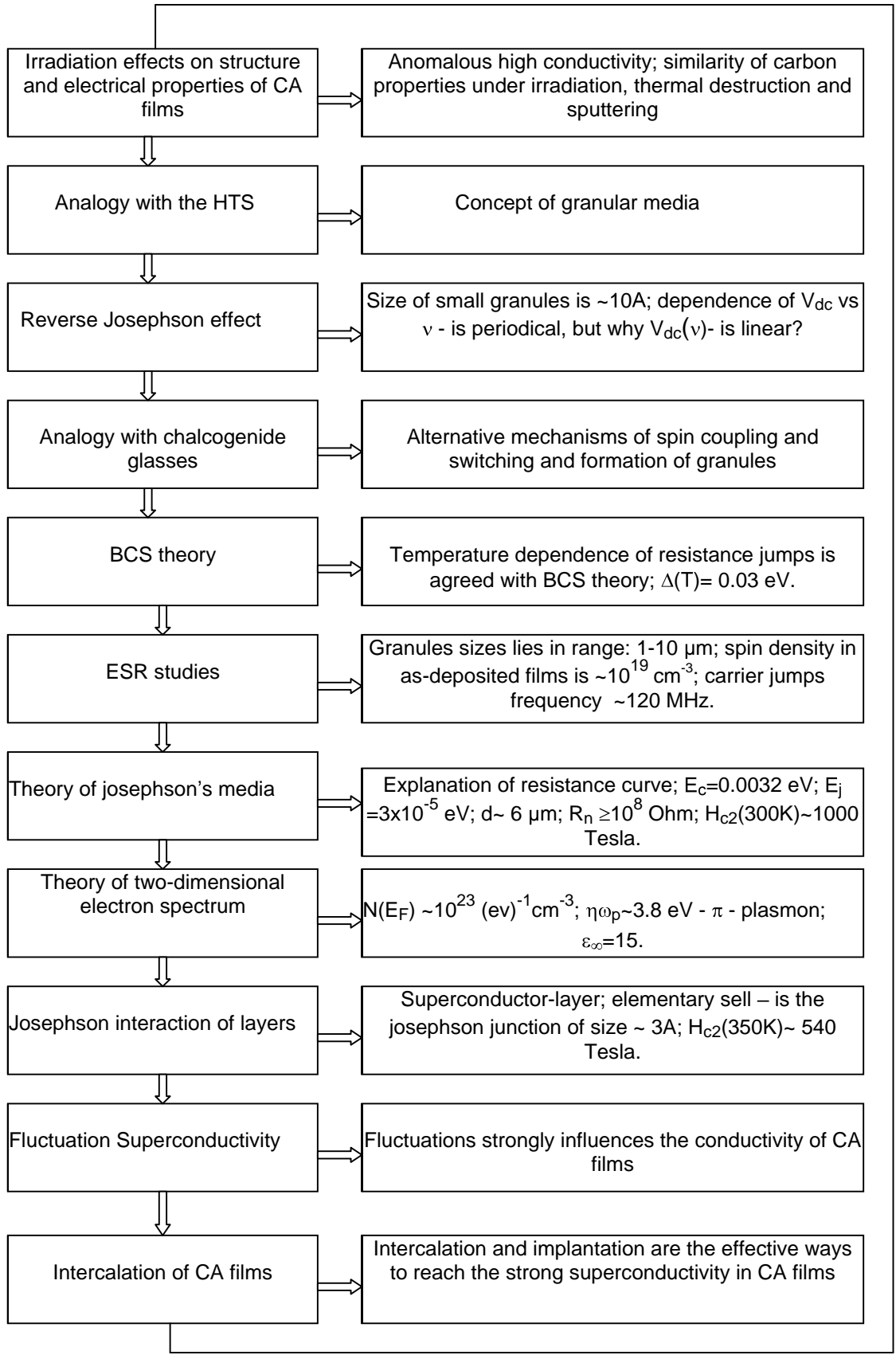

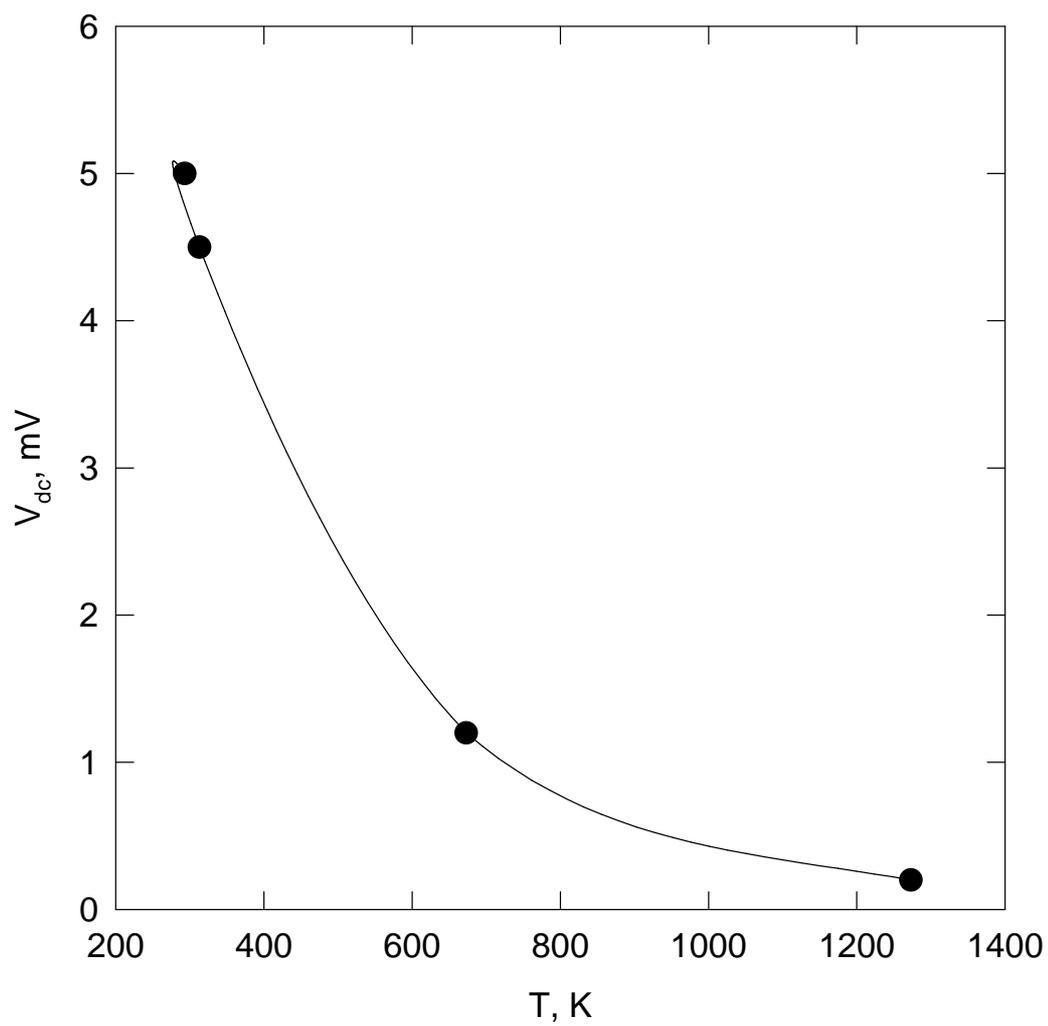

Fig.3. Temperature dependence of constant voltage $V_{dc}$ induced by microwave radiation in CA- films /10/

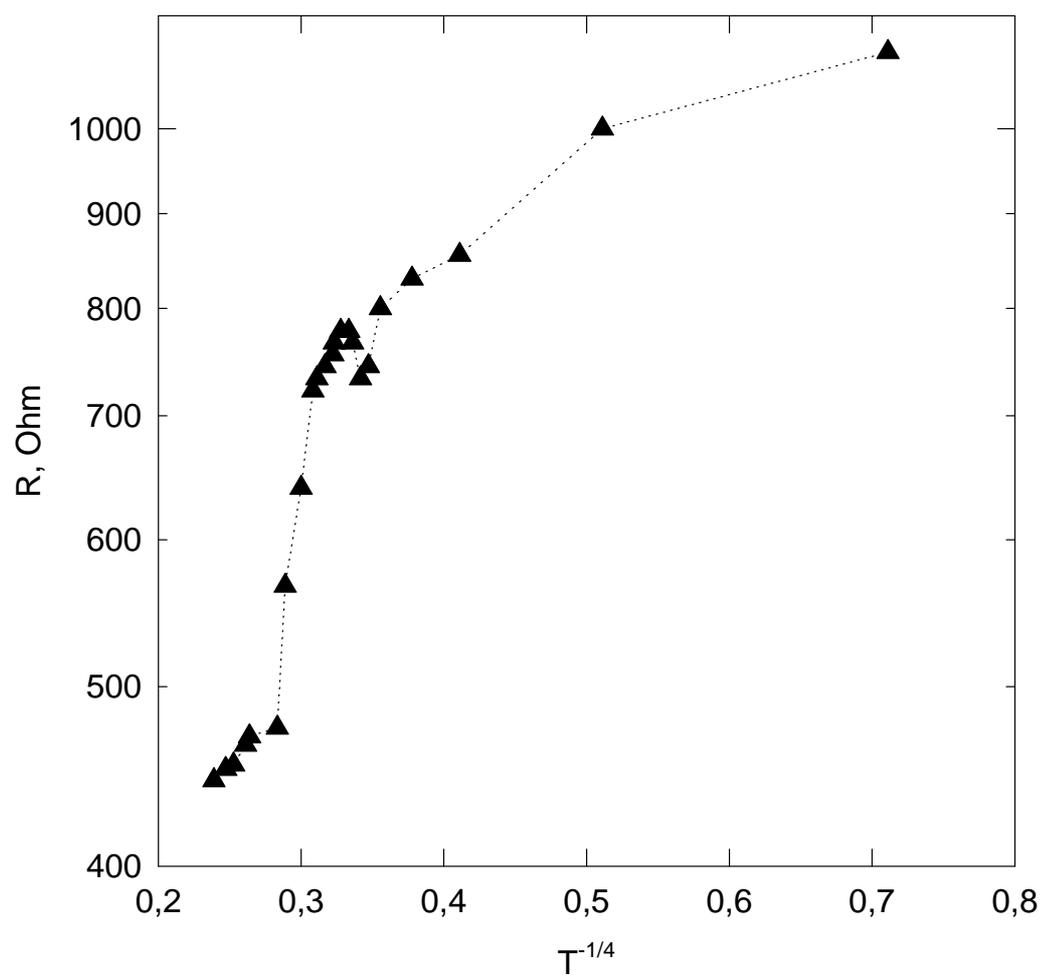

Fig.4. Resistance of CA – films as function of temperature /10/

Fig.5. Formation of pairs with alternating valence in chalcogenide glasses

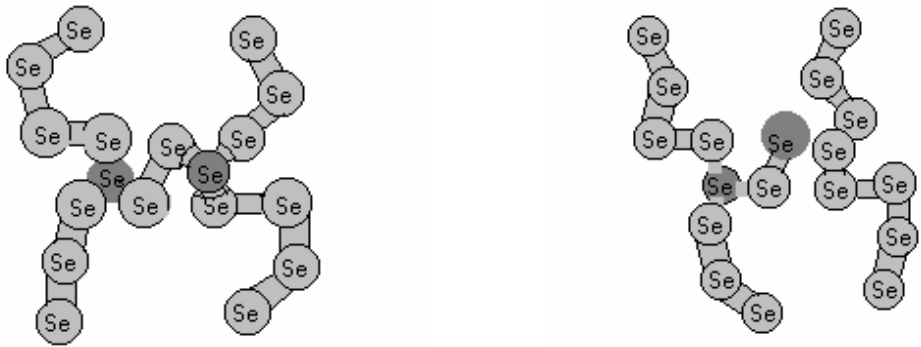

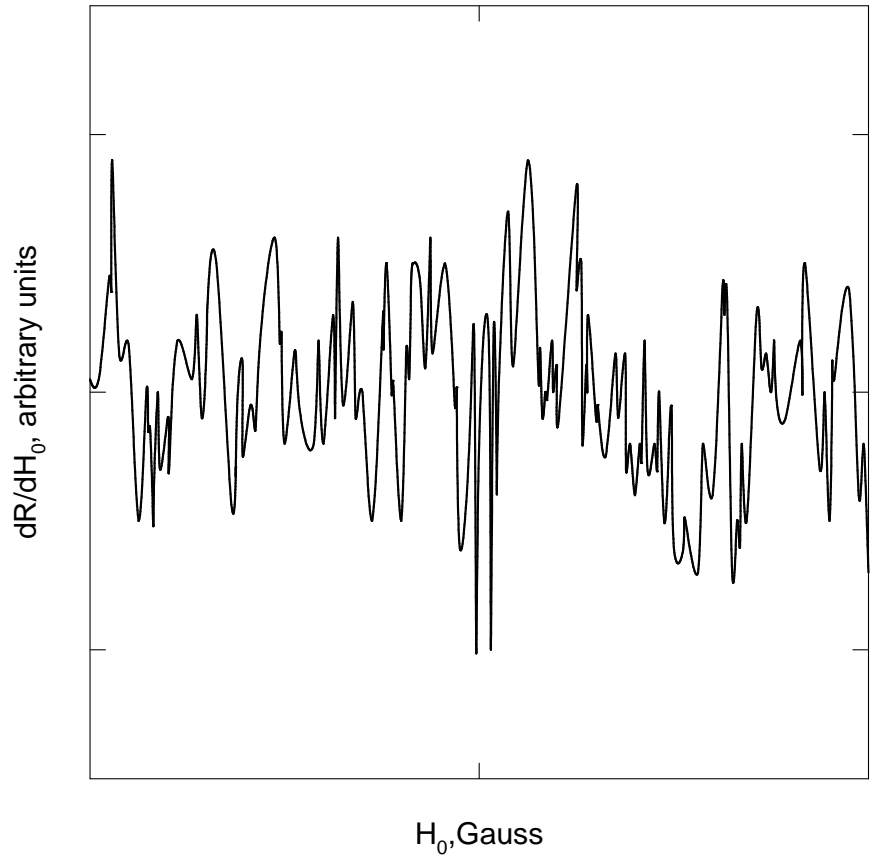

**Fig.6 Noise-like curve of microwave absorption in CA film**

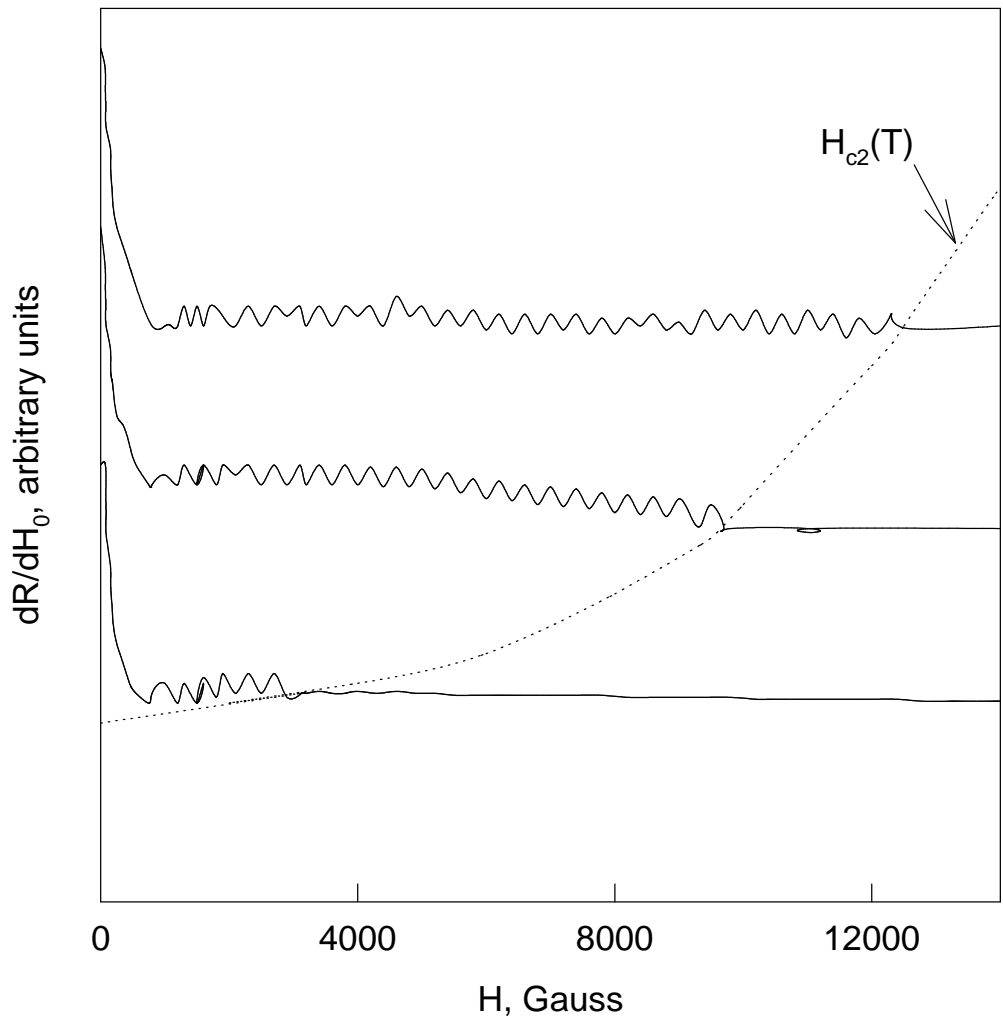

Fig.7. Microwave absorption in YBaCuO ceramic /95/. The noise-like signal isappears at $H>H_{C2}$

Table 1. Possible intercalants for graphite /109/.

| Intercalant | Sandwich thickness, A | Remarks |
|---|---|---|
| Li | 3.71 | Unstable in air |
| K | 5.35 | Unstable in air |
| Rb | 5.65 | Unstable in air |
| Cs | 5.94 | Unstable in air |
| $Br_2$ | 7.04 | Unstable in air |
| $HNO_3$ | 7.84 | Unstable in air |
| $HClO_3$ | 7.94 | Unstable in air |
| $SO_3$ | 7.96 | Unstable in air |
| $Cl_2O_7$ | 7.98 | Unstable in air |
| $AsF_5$ | 8.15 | Unstable in air |
| $SbF_5$ | 8.46 | Unstable in air |
| $FeCl_3$ | 9.37 | Stable in air |
| $CoCl_2$ | 9.40 | Unstable in air |
| $NiCl_2$ | 9.40 | Unstable in air |
| $SbCl_5$ | 9.42 | Stable in air |
| $AlCl_3$ | 9.54 | Unstable in air |
| KHg | 10.22 | Unstable in air |